\documentclass[aps,prc,twocolumn,superscriptaddress,amsmath,amssymb,showpacs,nofootinbib]{revtex4}
\usepackage{graphicx,color}

\begin{document}

\title{One fewer solution to the cosmological lithium problem}

\author{O.~S.~Kirsebom}
\email[Corresponding author: ]{oliverk@triumf.ca}
\affiliation{TRIUMF, Vancouver, BC V6T 2A3, Canada}
\affiliation{Department of Physics and Astronomy, Aarhus University, 8000 Aarhus C, Denmark}


\author{B.~Davids}
\affiliation{TRIUMF, Vancouver, BC V6T 2A3, Canada}

\begin{abstract}

Data from a recent $^9$Be($^3$He,$t)^9$B measurement are used to rule out a possible solution to the cosmological lithium problem based on conventional nuclear physics.

\end{abstract}


\maketitle


The primordial abundance of $^7$Li inferred from
observational data is roughly a factor of three below the abundance
predicted by the standard theory of big bang nucleosynthesis (BBN)~\cite{wagoner67}
using the baryon-to-photon ratio $\eta = 6.19(15)\times
10^{-10}$~\cite{wmap} determined mainly from measurements
of the cosmic microwave background radiation. 
In contrast, there is good agreement for $^2$H and $^4$He.
Taking into account the estimated uncertainties
on the observationally inferred and the theoretically deduced $^7$Li abundances,
the significance of the discrepancy is 4.2$\sigma$--5.3$\sigma$~\cite{cyburt08}.
This constitutes one of the important unresolved problems of
present-day astrophysics and is termed the {\it cosmological lithium
  problem}. Among other possibilities, the discrepancy could be due to new physics beyond
the Standard Model of particle physics~\cite{cyburt09_ii}, errors in the observationally
inferred primordial lithium abundance\footnote{Lithium may be
destroyed in metal-poor stars through diffusion and turbulent
mixing~\protect\cite{korn06}.}, or incomplete nuclear physics input for the BBN
calculations. The present paper addresses the last possibility.

In standard BBN theory, assuming $\eta = 6.19(15)\times 10^{-10}$, most
$^7$Li is produced in the form of $^7$Be. Only much
later, when the Universe has cooled sufficiently for nuclei and
electrons to combine into atoms, does $^7$Be decay to $^7$Li through
electron capture. The temperature range of $^7$Be production is $T
\simeq 0.3-0.6$~GK, where the main mechanism for $^7$Be production is
$^3$He$(\alpha,\gamma){}^7$Be while the main mechanism for $^7$Be
destruction is $^7$Be$(n,p){}^7$Li followed by $^7$Li$(p,\alpha){}^4$He.
The rates of these reactions as well as the
reactions that control the supply of neutrons, protons, $^3$He, and
$\alpha$ particles are known with better than 10\% precision at BBN
temperatures~\cite{cyburt04}, resulting in an uncertainty of only 13\%
on the calculated $^7$Li abundance~\cite{cyburt08}.

\begin{figure}
  \centering
  \includegraphics[width=0.99\linewidth, clip=true, trim= 100 570 40
  70]{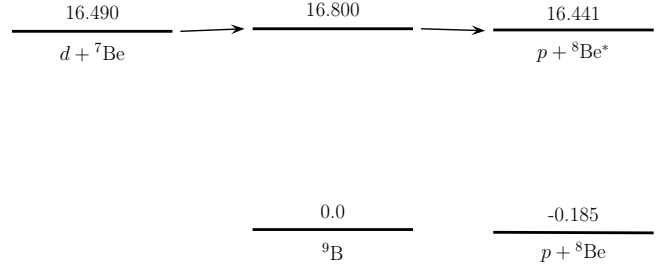}
  \caption{Schematic illustration of the proposed $^7$Be destruction
    mechanism, $d+{}^7$Be$\; \rightarrow {}^9$B$^{\ast} \rightarrow
    p+{}^8$Be$^{\ast}$. The energies are in MeV relative to the ground
    state of $^9$B. Subsequently, $^8$Be$^{\ast}$ breaks up into two
    $\alpha$ particles.}
  \label{fig : scheme}
\end{figure}

A recent theoretical paper~\cite{cyburt09} explores the possibility
of enhancing $^7$Be destruction through resonant reactions with $p$, $d$, $t$,
$^3$He, $\alpha$, leading to compound states in $^8$B, $^9$B,
$^{10}$B, $^{10}$C, $^{11}$C,
respectively. The paper concludes
that, of the known excited states in these isotopes~\cite{tilley04,
  ajzenberg-selove90}, only the 16.8~MeV state in $^9$B has the
potential to significantly influence $^7$Be
destruction\footnote{Ref.~\cite{chakraborty11} offers a more optimistic
  view, but only by adopting a somewhat flexible
  approach to basic principles of nuclear physics.}. 
(Note that in Ref.~\cite{cyburt09} this state is referred to as the 16.7 MeV state.)
The proposed destruction mechanism is shown schematically
in Fig.~\ref{fig : scheme}.
The 16.8~MeV state in $^9$B is formed by the fusion of $^7$Be with a
deuteron and decays by proton emission to a highly excited state in $^8$Be, 16.626~MeV
above the ground state, which subsequently breaks up into two $\alpha$
particles. (The last step is not shown in the figure.) The reason why
the decay must proceed by proton emission to the 16.626~MeV state in
$^8$Be and not e.g.\ the ground state will be explained below.

The reaction rate depends critically on the resonance energy, $E_r$,
i.e.\ the energy of the 16.8~MeV state relative to the $d+{}^7$Be threshold at
$S_d = 16.4901(10)$~MeV~\cite{tilley04}: if too far above the threshold, the tunneling process will
be too slow at BBN temperatures.
Furthermore, for the proposed destruction mechanism to be efficient,
the 16.8~MeV state must have an
appreciable width, $\Gamma_d$, for being formed in the $d+{}^7$Be
channel, but also an appreciable width, $\Gamma-\Gamma_d$,
for not decaying back to $d+{}^7$Be.
The energetically allowed decay modes competing with deuteron emission
are: $\gamma$, p, $\alpha$, and $^3$He. However, $\gamma$ and $^3$He
can safely be neglected.
A deuteron width, $\Gamma_d$, of the required magnitude can only be realized if
the 16.8~MeV state is not too close to the threshold.
The analysis of Ref.~\cite{cyburt09} shows that the cosmological lithium
problem can be resolved provided $E_r \simeq 170-220$~keV, $\Gamma_d
\simeq 10-40$~keV, and $\Gamma - \Gamma_d \simeq \Gamma_d$.
At the time Ref.~\cite{cyburt09} was written, the known properties of the
16.8~MeV state did not contradict these requirements: The 16.8~MeV state had
been observed in two experiments~\cite{pugh85, kadija87}. 
Its energy had been determined to be 16.7 MeV with an uncertainty of
100~keV, and only an upper limit of $100$~keV existed on its total
width. 
Its spin and parity had not been determined, though a tentative
$5/2^+$ assignment had been made~\cite{dixit91} based on comparison
to the mirror nucleus, $^9$Be. No information existed on its decay properties.

As noted in Ref.~\cite{cyburt09}, the simultaneous requirement of $E_r
\simeq 170-220$~keV and $\Gamma_d
\simeq 10-40$~keV is physically possible, but implies some rather
special properties for the 16.8~MeV state: a reduced deuteron
width comparable to the Wigner limit and a very large channel radius
of at least 9~fm.
In addition, the proposed destruction mechanism could only be reconciled
with the direct measurement of Ref.~\cite{angulo05} with considerable
difficulty: the proton and $\alpha$ decay of the 16.8~MeV state had
to be dominated by a single proton-decay branch to the 16.626~MeV,
$2^+$ state in $^8$Be, because decays to the lower-lying states in
$^8$Be would have produced protons of sufficient energy to be detected
by the experimental setup of Ref.~\cite{angulo05}.

Slightly above the 16.626~MeV state, at 16.922~MeV,
lies another $2^+$ state in $^8$Be. The two are nearly maximally mixed in
isospin~($I$)~\cite{hinterberger78} and are often referred to
as the {\it 2$^+$ doublet}. The structure of the 16.626~MeV state is
primarily that of a $1p_{1/2}$ proton orbiting a $^7$Li core in its
ground state, and the structure of the 16.922~MeV state is primarily
that of a $1p_{1/2}$ neutron orbiting a $^7$Be core in its ground
state~\cite{brentano90}. The analysis in
Ref.~\cite{dixit91} suggests that the structure of the 16.8~MeV state
in $^9$B is primarily that of a $2s_{1/2}$ proton orbiting the $I=1$
component of the $2^+$ doublet.  
As noted in Ref.~\cite{cyburt09}, this provided reason to think
that the overlap between the 16.8~MeV state in $^9$B and
$p+{}^8$Be$^{\ast}$ might indeed be considerably larger for the
16.626~MeV state than for any of the lower-lying states in $^8$Be.

The nuclear physics input for BBN calculations has recently been
reexamined in Ref.~\cite{boyd10}, which includes new reactions, studies the
potential effects of reactions for which data do not exist, studies
the effects of non-thermal particles (highly energetic particles
produced in exothermic reactions), in particular neutrons which take
much longer to thermalize than charged particles, and includes
thermal excitation of the first excited states in $^7$Li and
$^7$Be. Ref.~\cite{boyd10} concludes that there is little chance of solving the
cosmological lithium problem with conventional nuclear physics,
but retains the destruction mechanism proposed by Ref.~\cite{cyburt09} as an
``alluring'' possibility.

A very recent paper~\cite{scholl11} reports on a new
$^9$Be$(^3$He,$t)^9$B measurement performed with the purpose of studying Gamow-Teller
transition strengths in the $A=9$ system. The beam energy was
140~MeV/nucleon, and tritons were detected in a high-resolution
spectrometer at scattering angles around $0^{\circ}$. The excitation energy
resolution achieved was 30~keV. The 16.8~MeV state is strongly excited,
and its energy and width are determined to be 16.800(10)~MeV and
81(5)~keV, respectively, in good agreement with the two previous experiments.
The nearby $J^{\pi} = 1/2^{-}$, $I=3/2$ state at 17.076(4)~MeV is also
strongly excited and its energy is well-known from its $\gamma$ decay
to the ground state. This gives strong confidence in the new energy
determination. It is also worth noting that the observed angular
distribution of the 16.8~MeV state is consistent with the proposed
$5/2^+$ assignment.

To assess the consequence of the new experimental data for the proposed
destruction mechanism, we employ the standard Kawano/Wagoner BBN
code~\cite{kawano92,smith93}. 
We modify the $^7$Be($d$,$p$) reaction rate by adding the extra
term~\cite{clayton68}
\begin{align}\label{eq : extra term}
N_A \langle \sigma v \rangle \; &= \; N_A \left( 8/ \pi \mu_{27} \right)^{1/2} 
(kT)^{-3/2} \nonumber \\
&\phantom{xx} \times \int_0^{\infty} E \; \sigma(E) \; \exp{ ( -E/kT ) } \;
\textrm{d}E \; ,
\end{align}
where $N_A$ is Avogadro's constant, $\mu_{27}$ is the reduced mass, $k$ is Boltzmann's
constant, $T$ is the temperature, $E$ is the relative kinetic
energy, and $\sigma(E)$ is the cross section for $d+{}^7$Be$\; \rightarrow {}^9$B$^{\ast} \rightarrow
    p+{}^8$Be$^{\ast}$, given by the single-channel, single-level approximation of the
$R$-matrix theory~\cite{lane58}:
\begin{align*}
\sigma(E)  \; &= \; \pi \lambdabar^2 \; \omega \; \frac{ \Gamma_d
  (\Gamma - \Gamma_d) }{ (E-E_r-\Delta)^2 + (\Gamma/2)^2 } \; ,
\end{align*}
where $\lambdabar = \hbar / p = \hbar / ( 2\mu_{27} E )^{1/2}$, and $\omega$
is a statistical weight factor that depends on the spins involved,
\begin{equation*}
\omega = \frac {2J+1} { (2j_1+1) (2j_2+1) } 
= \frac {2\times \tfrac{5}{2}+1} { (2 \times 1+1) (2 \times \tfrac{3}{2}
  +1) } = 0.5 \; ,
\end{equation*}
where $J=5/2$ is the (assumed) spin of the 16.8~MeV state, $j_1=1$ is
the spin of the deuteron, and $j_2=3/2$ is the spin of $^7$Be.
Furthermore, $\Gamma = \Gamma_{\gamma} + \Gamma_p + \Gamma_d +
\Gamma_{^3\textrm{He}} + \Gamma_{\alpha}$ is the total
width. We assume $\Gamma_{\gamma}$, $\Gamma_{^3\textrm{He}}$,
and $\Gamma_{\alpha}$ to be negligible and $\Gamma_p$ to
be dominated by the decay to the 16.626~MeV state in $^8$Be. 
As the 16.8~MeV state is located close to threshold, the energy
dependence of the deuteron width must be taken into account~\cite{teichmann52}:
\begin{equation*}
\Gamma_d = 2 \; P_{\ell=1}(E) \; \gamma_d^2 \; .
\end{equation*}
Similarly, for the proton width:
\begin{equation}\label{eq : proton width}
\Gamma_p = 2 \; P_{\ell=0}(E^{\prime}) \; \gamma_p^2 \; ,
\end{equation}
where $P_{\ell}$ is the penetrability, $\ell$ is the
orbital angular momentum, $\gamma_d$ ($\gamma_p$) is the
deuteron (proton) reduced width, and $E^{\prime}$ is
the $p+{}^8$Be$^{\ast}$ relative kinetic energy, 
\begin{equation*}
E^{\prime} = E + S_d - S_p - 16.626\textrm{ MeV} 
\end{equation*}
with $S_p=-0.1851(10)$~MeV \cite{tilley04}. 
We note that Eq.~(\ref{eq : proton width}) is only approximately
valid, as it assumes that the width of the 16.626~MeV state in $^8$Be
can be neglected, whereas the state actually has a considerable width of
108.1(5)~keV~\cite{tilley04} with an asymmetric line shape owing to
interference with the 16.922~MeV state. Still, the approximation is
adequate for the present analysis.
Finally, the shift, $\Delta$, is calculated as
\begin{equation*}
\Delta \; = \; - \left( \, S_{\ell=1}(E) - B \, \right) \gamma_d^2 \; - \left( \,
S_{\ell=0}(E^{\prime}) - B^{\prime} \, \right) \gamma_p^2\; ,
\end{equation*}
where $S_{\ell}$ is the shift function, and the boundary conditions
are $B=S_{\ell=1}(E_r)$ and $B^{\prime}=S_{\ell=0}(E_r^{\prime})$. 
The definitions of $P_{\ell}$ and $S_{\ell}$ are given in Ref.~\cite{lane58}.
To evaluate $P_{\ell}$ and $S_{\ell}$, suitable channel
radii, $a_{27}$ and $a_{18}$, must be chosen for the formation and destruction
channel. 

\begin{figure}[tb]
  \centering
  \includegraphics[width=0.9\linewidth,clip=true,trim= 10 55 70
  95]{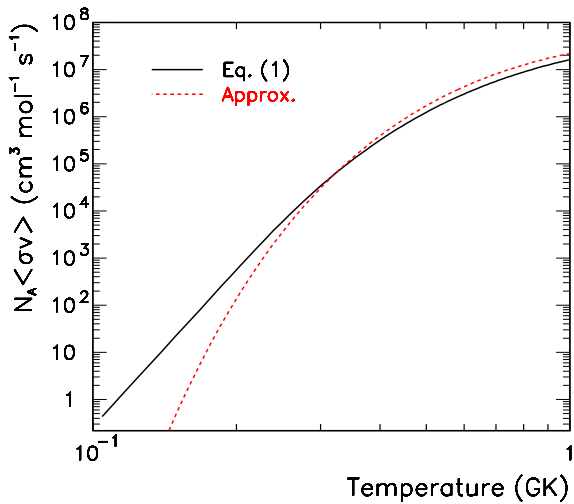}
\caption{Maximal resonant contribution of
  the 16.8 MeV state in $^9$B to the $^7$Be$(d,p)$ reaction rate
  calculated from Eq.~(\protect\ref{eq : extra term}) versus
  temperature. For comparison, the rate calculated
  using the narrow resonance approximation is also shown.}
\label{fig : rate}
\end{figure}

Relying on the data from the new
$^9$Be($^3$He,$t)^9$B measurement~\cite{scholl11}, we use
$E_r=310(10)$~keV for the resonance energy and $\Gamma^0 = 81(5)$~keV
for the total width. The superscript 0 refers to the value at
resonance energy, i.e.\ at $E=E_r$.
To maximize the reaction rate, we chose $\gamma_d$ and $\gamma_p$ such
that $\Gamma_d^0 = \Gamma_p^0 = 0.5 \, \Gamma^0$.
We do not have complete liberty in our choice of $\gamma_d$ and
$\gamma_p$ as they should not exceed the corresponding Wigner
limits, $\gamma_{\textrm{W},d}^2 = 3\hbar^2/(2\mu_{27} a_{27}^2)$ and
$\gamma_{\textrm{W},p}^2 = 3\hbar^2/(2\mu_{18} a_{18}^2)$.
We find that a permissible choice of $\gamma_d$ only exists for
$a_{27} > 6.5$~fm, whereas a permissible choice of $\gamma_p$ exists
for essentially any value of $a_{18}$. 
For the present calculation, we chose $a_{27} = 7$~fm and $a_{18}=5$~fm,
yielding $\gamma_d^2 / \gamma_{\textrm{W},d}^2 = 0.93$ and $\gamma_p^2
/ \gamma_{\textrm{W},p}^2 = 0.09$. As argued in
Ref.~\protect\cite{cyburt09}, $a_{27}=7$~fm represents a physically
plausible choice of channel radius.
The temperature dependence of the reaction rate is shown in
Fig.~\ref{fig : rate}. 
For comparison, we also show the rate obtained
in the narrow resonance approximation. 
We find that the reduction in $^7$Li abundance caused by the 
inclusion of the resonant contribution of the 16.8 MeV state in $^9$B 
to the $^7$Be$(d,p)$ reaction rate is at most 3.5(8)\%. 
This result is essentially independent of the choice of
channel radii. 
The quoted uncertainty mainly reflects the 10~keV uncertainty on the energy
determination of the 16.8~MeV state with a small contribution (0.2\%) from the
5~keV uncertainty on the width determination. 

We stress that the assumption of a dominant proton-decay branch to the
16.626~MeV state is,
by no means, important to the conclusion of the present analysis. If
the decay is assumed to proceed by proton emission to lower-lying
states in $^8$Be or $\alpha$ emission to $^5$Li, a similar reduction
in $^7$Li abundance is obtained.
The assumption of a dominant proton-decay branch to the 16.626~MeV
state was
made mainly to avoid conflict with the direct measurement of Ref.~\cite{angulo05}.

In summary, we have shown that the 16.8~MeV state in $^9$B is unable
to enhance the $^7$Be$(d,p)$ reaction rate by the amount needed to resolve the
cosmological lithium problem. With the new precise determination of the
energy of the 16.8~MeV state~\cite{scholl11}, the reduction in
$^7$Li abundance owing to the inclusion of the resonant contribution
of the 16.8 MeV state to the $^7$Be$(d,p)$ reaction rate is at
most 3.5(8)\% and probably much lower depending on the decay
properties of the 16.8~MeV state which remain unknown.  In line with
Ref.~\cite{boyd10}, we conclude that all possibilities for solving the
cosmological lithium problem by conventional nuclear physics means now
seem to have been exhausted.\\

OSK acknowledges support from the Villum Kann Rasmussen Foundation.
BD acknowledges support from the Natural Sciences and Engineering
Research Council of Canada. 
TRIUMF receives federal funding via a contribution agreement through
the National Research Council of Canada.


\begin{thebibliography}{99}

\bibitem{wagoner67} R.\ V.\ Wagoner, W.\ A.\ Fowler, and F.\ Hoyle,
  Astrophys.\ J.\ {\bf 148}, 3 (1967).

\bibitem{wmap} E.\ Komatsu {\it et al.}\ (WMAP Collab.),
  Astrophys.\ J.\ Suppl.\ {\bf 192}, 18 (2011).

\bibitem{cyburt08} R.\ H.\ Cyburt {\it et al.}, J.\ Cos.\ Astropart.\
  Phys.\ 11, 012 (2008).

\bibitem{cyburt09_ii} For an example, see R.~H.~Cyburt {\it et al.},
  J.\ Cos.\ Astropart.\ Phys.\ 10, 021 (2009). 

\bibitem{korn06} A.\ J.\ Korn {\it et al.}, Nature {\bf 442}, 657 (2006).

\bibitem{cyburt04} R.\ H.\ Cyburt, Phys.\ Rev.\ D {\bf 70}, 023505 (2004).

\bibitem{cyburt09} R.\ H.\ Cyburt and M.\ Pospelov, arXiv:0906.4373v1
[astro-ph] (2009).

\bibitem{chakraborty11} N.\ Chakraborty, B.\ D.\ Fields, and K.\ A.\
  Olive, Phys.\ Rev.\ D {\bf 83}, 063006 (2011).

\bibitem{tilley04} D.\ R.\ Tilley {\it et al.}, Nucl.\ Phys.\ A {\bf 745}, 155 (2004).
\bibitem{ajzenberg-selove90} F.\ Ajzenberg-Selove, Nucl.\ Phys.\ A {\bf 506}, 1 (1990).

\bibitem{pugh85} B.\ Pugh, Ph.D.\ thesis, Massachusetts Institute of
Technology (1985).

\bibitem{kadija87} K.\ Kadija, G.\ Pai{\'c}, B.\ Antolkovi{\'c}, A.\ Djaloeis, and J.\ Bojowald, Phys.\ Rev.\ C {\bf 36}, 1269
(1987).

\bibitem{dixit91} S.\ Dixit {\it et al.}, Phys.\ Rev.\ C {\bf 43}, 1758 (1991).

\bibitem{angulo05} C.\ Angulo {\it et al.}, Astrophys.\ J.\ {\bf 630}, L105 (2005).

\bibitem{hinterberger78} F.\ Hinterberger {\it et al.}, Nucl.\ Phys.\
  A {\bf 299}, 397 (1978).

\bibitem{brentano90} P.\ von Brentano, Phys.\ Lett.\ B {\bf 246}, 320 (1990).

\bibitem{boyd10} R.\ N.\ Boyd, C.\ R.\ Brune, G.\ M.\ Fuller, and C.\ J.\ Smith, Phys.\ Rev.\ D {\bf 82}, 105005 (2010).

\bibitem{scholl11} C.\ Scholl {\it et al.}, Phys.\ Rev.\ C {\bf 84}, 014308 (2011).

\bibitem{kawano92} L.\ H.\ Kawano, Report No.\ FERMILAB-Pub-92/04-A,
  preprint, 1992.
\bibitem{smith93} M.\ S.\ Smith, L.\ H.\ Kawano, and R.\ A.\ Malaney,
  Astrophys.\ J.\ Suppl.\ Ser.\ {\bf 85}, 219 (1993).

\bibitem{clayton68} D.\ D.\ Clayton, Principles of Stellar Evolution
  and Nucleosynthesis, New York, McGraw-Hill, 1968.

\bibitem{lane58} A.\ M.\ Lane and R.\ G.\ Thomas, Rev.\ Mod.\ Phys.\ {\bf 30}, 257 (1958).

\bibitem{teichmann52} T.\ Teichmann and E.\ P.\ Wigner, Phys.\ Rev.\ {\bf 87}, 123 (1952).

\end{thebibliography}
\end{document}